\documentstyle[pra,aps,twocolumn,epsf,floats]{revtex}
\def\geqap{\,\raise 2pt \hbox{$>\kern-11pt \lower 5pt \hbox{$\sim$}$}\,}
\def\leqap{\,\raise 2pt \hbox{$<\kern-10pt \lower 5pt \hbox{$\sim$}$}\,}
\begin{document}
\draft
\twocolumn[\hsize\textwidth\columnwidth\hsize\csname @twocolumnfalse\endcsname

\title{Transition between two ferromagnetic states driven by orbital 
ordering in
La$_{0.88}$Sr$_{0.12}$MnO$_{3}$}
\author{Y.Endoh$^{1,5}$, K.Hirota$^{1,5}$, S.Ishihara$^{2,5}$,
S.Okamoto$^{2,5}$, Y.Murakami$^{3,5}$, A.Nishizawa$^{1}$,
T.Fukuda$^{4,5}$, H.Kimura$^{1,5}$,
H.Nojiri$^{2,5}$, K.Kaneko$^{2,5}$, and S.Maekawa$^{2,5}$}
\address{$^1$Department of Physics, Tohoku University,  Sendai, 980-8578 
Japan}
\address{$^2$Institute for Materials Research, Tohoku University,  Sendai,
980-8577 Japan}
\address{$^3$Photon Factory, Institute of Materials Structure Science,  KEK,
Tsukuba 305-0801, Japan}
\address{$^4$Department of Synchrotron Radiation Facility Project, JAERI, 
Sayo
679-5143, Japan}
\address{$^5$CREST, Japan Science and Technology Corp., Tsukuba 305-0047, 
Japan}

\date{\today}
\maketitle
\begin{abstract} 
A lightly doped perovskite mangantite
La$_{0.88}$Sr$_{0.12}$MnO$_{3}$ exhibits a phase transition at $T_{OO}=145$~
K 
from a ferromagnetic metal ($T_{C}=172$~K) to a novel ferromagnetic 
insulator.  
We identify that the key parameter in the transition is the orbital degree 
of freedom in $e_g$ electrons.  By utilizing the resonant x-ray scattering
technique, orbital ordering is directly detected below $T_{OO}$, in spite
of a significant diminution of the cooperative Jahn-Teller distortion.  
The new experimental features are well described by a theory treating
the orbital degree of freedom under strong electron correlation. 
The present experimental and theoretical studies  uncover a crucial role 
of the orbital degree in the metal-insulator transition in lightly doped
manganites. 
\end{abstract}
]

\narrowtext

Colossal magnetoresistance (CMR), recently discovered in perovskite 
manganites,
occurs in the vicinity of metal-insulator (MI) transition. It was proposed
many years ago that the double-exchange (DE) mechanism plays an essential 
role
to realize the ferromagnetic metallic state in doped
manganites \cite{zener,anderson}.  However, the CMR effects cannot be 
explained
within this simple concept\cite{millis} and additional ingredients, such as
lattice distortion, electron correlation, and orbital degree of freedom, 
are stressed.  To reveal the mechanism of the MI transition and its mutual
relation to CMR, it is essential to study the lightly doped regime in 
detail,
where several phase boundaries are entangled.

In La$_{1-x}$Sr$_{x}$MnO$_{3}$ around $x \sim 0.12$, the temperature
dependence of electrical resistivity shows metallic behavior below
$T_{C}$ consistent with the DE picture.  As temperature decreases further,
however, it shows a sharp upturn below a certain 
temperature \cite{urushibara}, 
defined $T_{OO}$ in the present paper.  Note that a transition from
ferromagnetic metallic (FM) state to the ferromagnetic insulating (FI) state
occurs at $T_{OO}$.  Kawano {\it et al.}\cite{kawano} revealed by
neutron diffraction that La$_{0.875}$Sr$_{0.125}$MnO$_{3}$
($T_{C}=230$~K) exhibits successive structural phase transitions;
high-temperature pseudo-cubic phase (O$^{*}$: $a \sim b \sim c/\sqrt{2}$) to
intermediate Jahn-Teller (JT) distorted orthorhombic phase (O$'$: $b > a \gg
c/\sqrt{2}$) at $T_{H}=260$~K and to low-temperature O$^{*}$ phase at
$T_{OO}=160$~K. Here, we use orthorhombic
$Pbnm$ notation.  These complicated behaviors are far beyond the simple DE
scenario.

In this Letter, we report that the MI transition in
La$_{0.88}$Sr$_{0.12}$MnO$_{3}$ is actually driven by orbital ordering, 
which
was directly observed by the recently developed decisive technique, i.e., 
the
resonant x-ray scattering\cite{murakami,ishihara3}. It is counter-intuitive 
that
the orbital ordering appears in the FI phase where a long-range cooperative 
JT
distortion significantly diminishes \cite{argyriou}.  As discussed later, 
this
orbital ordering can be realized by the super-exchange (SE) process under 
strong
electron correlation together with ferromagnetic ordering, not necessarily
associated with a cooperative JT distortion.  This ferromagnetic SE 
interaction
coexists with the DE interaction which facilitates carrier mobility above
$T_{OO}$.  The transition from FM to FI can be induced by applying
a magnetic field \cite{senis,uhlenbruck,nojiri}. 
Our theoretical calculation, where the orbital degree and
the electron correlation are considered, well reproduces these
unconventional experimental results in La$_{0.88}$Sr$_{0.12}$MnO$_{3}$.

\begin{figure}
\epsfxsize=6.5cm
\centerline{\epsffile{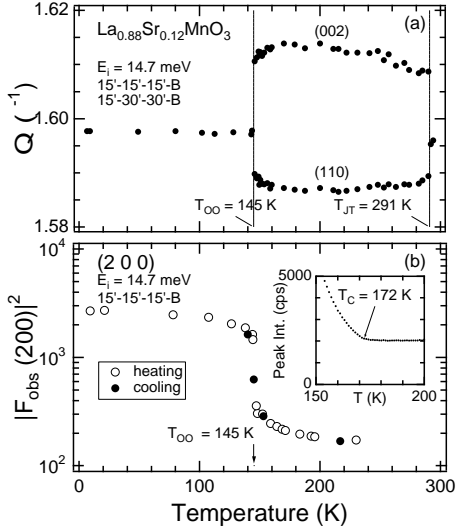}}
%
\caption{Temperature dependence of (a) lattice parameter, and integrated
intensities of (b) $(2\ 0\ 0)$ ferromagnetic Bragg reflection measured with
14.7~meV neutrons.}
\label{fig1}
\end{figure}

We have grown series of single crystals by the floating-zone method using a 
lamp
image furnace.  Typical mosaicness measured by neutron diffraction is less 
than
0.3$^{\circ}$ FWHM, indicating that the samples are highly crystalline.  We 
have
carried out neutron-diffraction of an $x=0.12$ single crystal using the 
TOPAN
triple-axis spectrometer in the JRR-3M reactor in JAERI.  
As shown in Fig.~\ref{fig1},  
we found successive structural phase transition and magnetic
ordering consistent with Ref.~\onlinecite{kawano}, though transition
temperatures are different reflecting a slight discrepancy in the hole
concentration, i.e., $x=0.12$ and $0.125$. 
However, high-resolution synchrotron x-
ray
powder diffraction by Cox {\it et al.}\cite{cox} on a carefully crushed 
small
crystal from the same batch reveals that the intermediate phase is 
monoclinic
and that the low-temperature phase is triclinic though the distortion is
extremely small. An electron-diffraction study by 
Tsuda {\it et al.} \cite{tsuda}
using the same batch gave consistent results.  Note that typical x-ray peak
widths are $\sim 0.01^{\circ}$, close to the instrumental resolution.  This
indicates that the $d$-spacing distribution is negligible, which is another
proof of high quality of our samples.  

%
%
%
%

Here, we briefly mention the charge ordering proposed by Yamada {\it
et al}\cite{yamada}.  We have indeed confirmed the superlattices below
$T_{OO}$ by neutron and x-ray scatterings.  In the x-ray study, however, the
energy dependence of the superlattice peak around the Mn K-edge does not 
show a
resonance feature\cite{inami,fukuda} which is a characteristic in
Mn$^{3+}$/Mn$^{4+}$ charge ordering\cite{murakami}.  We thus conclude that, 
below
$T_{OO}$, there appears a long-range structural modulation along the $c$-
axis
though neither a conventional charge ordering nor a long-range cooperative 
JT
distortion as seen in the O$'$ phase exist.

Ferromagnetic ordering below $T_{C}=172$~K was observed by neutron 
diffraction as shown in Fig.~\ref{fig1}(b).  With further decreasing 
temperature,
the $(2\ 0\ 0)$ peak exhibits a discontinuous increase at temperature
corresponding to $T_{OO}$, where the structural phase transition shown in
Fig.~1(a) occurs.  We found that magnetic Bragg peaks appear at $(0\ 0\ l)$
($l={\rm odd}$) below $T_{OO}$ indicating antiferromagnetic component.  
When the magnetic structure below $T_{OO}$ is interpreted 
as a canted structure, the canting angle is, however, small and the
FI state is a good approximation.

The orbital states were observed by synchrotron x-ray diffraction
measurements on four-circle spectrometers at beamlines 4C and 16A2 in the
Photon Factory in KEK. We have tuned 
the
incident energy near the Mn K- edge (6.552~KeV).  The $(0\ 1\ 0)$ plane of a
La$_{0.88}$Sr$_{0.12}$MnO$_{3}$ single crystal ($\sim 2\phi \times 2$~mm)
from the same batch, which was carefully polished, was
aligned within the scattering plane. 

\begin{figure}
\epsfxsize=7.0cm
\centerline{\epsffile{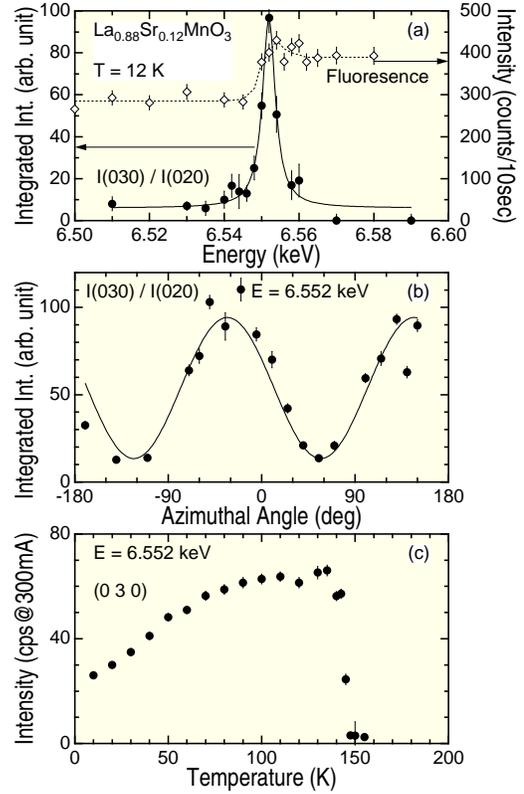}}
%
%
\caption{(a) Energy dependence of intensity of the orbital-ordering 
reflection
$(0\ 3\ 0)$ at $T=12$~K. The
dashed curve represents fluorescence showing the resonant energy (6.552~KeV)
corresponding to the Mn K edge. (b) The azimuthal angle dependence
of $(0\ 3\ 0)$. The solid line is two-fold squared sine curve of angular
dependence. (c) Temperature dependence of $(0\ 3\ 0)$ peak intensity.}
\label{fig2}
\end{figure}

Figure~\ref{fig2}(a) shows the incident energy dependence of $(0\ 3\ 0)$
peak, which is structurally forbidden, at 12~K.  The peak exhibits a sharp
enhancement at the Mn K-edge, determined experimentally from
fluorescence.  As discussed in La$_{1.5}$Sr$_{0.5}$MnO$_{4}$ and also
in LaMnO$_{3}$\cite{murakami}, the appearance of such a forbidden peak is
considered as a signature of orbital ordering of Mn$^{3+}$ $e_{g}$ 
electrons:
Orbital ordering gives rise to anisotropy in the anomalous scattering 
factor,
which is  enhanced and thus visible near the Mn K-edge.  The antiferro(AF)-type 
orbital 
ordering is
directly confirmed by rotating the crystal around the scattering
vector kept at $(0\ 3\ 0)$, i.e., azimuthal scan. 
Fig.~\ref{fig2}(b) shows the azimuthal scan, clearly revealing square of
sinusoidal angle dependence of two-fold symmetry. 
%
%
%
%

Although the orbital ordering in La$_{0.88}$Sr$_{0.12}$MnO$_{3}$ seems
similar to that of LaMnO$_{3}$, there is a marked difference.  As shown in
Fig.~\ref{fig2}(c), the orbital ordering appears only below $T_{OO}=145$~K, 
where
the cooperative JT distortion disappears or significantly decreases.  On the
other hand, in LaMnO$_{3}$ where the orbital ordering appears in the JT 
distorted orthorhombic phase\cite{murakami}, it has been believed that long-
range
arrangement of JT distorted MnO$_{6}$ octahedra facilitates
$d_{3x^2-r^2}/d_{3y^2-r^2}$ type orbital ordering.  The orbital ordering is
consistent with the spin-wave dispersion reported by Hirota {\it et
al.}\cite{hirota1}, who proposed the dimensional crossover in lightly doped
La$_{1-x}$Sr$_{x}$MnO$_{3}$\cite{hirota2}. The spin dynamics of
La$_{1-x}$Sr$_{x}$MnO$_{3}$ drastically changes from two-dimenional state as
seen in LaMnO$_{3}$, due to the AF-type orbital ordering of
$d_{3x^2-r^2}/d_{3y^2-r^2}$, to three-dimensional isotropic ferromagnetic 
state
around $x \approx 0.1$.   Therefore, we anticipate that
La$_{0.88}$Sr$_{0.12}$MnO$_{3}$ should have a different orbital state, e.g.,
 the
hybridization of $d_{z^2-x^2(y^2-z^2)}$ and $ d_{3x^2-r^2(3y^2-r^2)}$.   
Note
that the intensity of $(0\ 3\ 0)$ resonant peak is significantly reduced at 
low
temperatures compared with that just below $T_{OO}$, indicating that the AF-
type
orbital ordering becomes reduced with decreasing temperature.  This 
reduction is
not necessarily due to the instability of orbital ordering at low 
temperatures
because the gradual change of type of the orbital ordering , e.g., AF-type to
ferro-type, gives rise to the effect.

Now we theoretically reveal the microscopic mechanism of the newly found
experimental features.  The spin and orbital states are investigated by
utilizing the model Hamiltonian where the 
spin and orbital degrees of freedom are
treated on an equal footing together with the strong electron correlation \cite{ishihara1}: 
$
{\cal H}={\cal H}_{t}+{\cal H}_{J}+{\cal H}_{H}+{\cal H}_{AF}
$.
The first and second terms correspond to the so-called
$t$- and $J$-terms in the $tJ$-model for $e_g$ electrons.  These
are given by \cite{ham}
$
{\cal H}_t=\sum_{\langle i j \rangle \gamma \gamma' \sigma }
t_{ij}^{\gamma \gamma'} \tilde d_{i \gamma \sigma}^\dagger 
\tilde d_{j \gamma' \sigma} + H.c.  
$
and
\begin{eqnarray}
{\cal H}_{J}=&-&2J_1\sum_{\langle ij \rangle } 
\Bigl ( {3 \over 4} n_i n_j + \vec S_i \cdot \vec S_j   \Bigr )
\Bigl ( {1 \over 4}  - \tau_i^l \tau_j^l \Bigr ) \nonumber \\
     &-&2J_2\sum_{\langle ij \rangle } 
\Bigl ( {1 \over 4} n_i n_j  - \vec S_i \cdot \vec S_j   \Bigr )
\Bigl ( {3 \over 4}   + \tau_i^l \tau_j^l +\tau_i^l+\tau_j^l \Bigr ) \ , 
\label{eq:hj}
\end{eqnarray}
where
$
\tau_i^l =\cos({2 \pi \over 3} n_l)
T_{iz}-\sin ({2 \pi \over 3} n_l) T_{ix}  
$ and 
$(n_x,n_y,n_z)$$=(1,2,3)$. 
$l$ denotes a direction of a bond connecting $i$ and $j$ sites.  
$\tilde d_{i \gamma \sigma}$ is the annihilation operator of $e_g$ electron at site $
i$
with spin $\sigma$ and orbital $\gamma$ with excluding double 
occupancy. 
The spin and orbital states are denoted by the spin operator $\vec S_i$ and
the pseudo-spin operator $\vec T_i$, respectively.  The latter describes which of the 
orbitals
is occupied.  The third and fourth terms  in the Hamiltonian describe the 
Hund
coupling: ${\cal H}_{H}=-J_H\sum_i \vec S_{t i} \cdot \vec S_i$ and the AF
magnetic interaction between $t_{2g}$ spins: 
${\cal H}_{AF}=J_{AF}\sum_{\langle ij \rangle} \vec S_{t i} \cdot \vec S_{t 
j}$, 
respectively,  where $\vec S_{t i}$ is the spin operator for 
$t_{2g}$ electrons with $S=3/2$. 
Since the cooperative JT distortion has been 
experimentally found to be weak around $x \sim 0.1$, 
the electron-lattice coupling is neglected in the model. 
As seen in the first term of ${\cal H}_{J}$, 
the ferromagnetic SE interaction 
results from the orbital degeneracy and 
the Hund coupling between $e_g$ electrons \cite{goodenough,kanamori,kugel}:
Through the coupling between spin and orbital degrees 
in ${\cal H}_{J}$, the ferromagnetic ordering and AF-type 
orbital ordering are cooperatively stabilized.  

%
%
\begin{figure}
\epsfxsize=6.5cm
\centerline{\epsffile{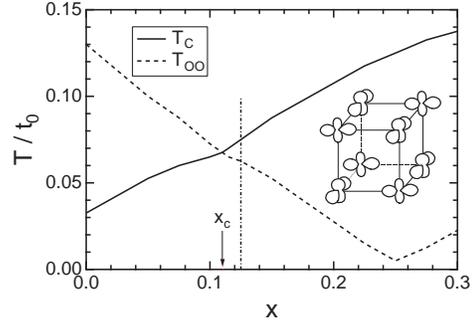}}
%
%
\caption{The spin and orbital phase diagram as a function  of 
carrier concentration $(x)$ and temperature $(T)$.  The straight and 
broken
lines are for the spin and orbital ordering temperatures.  
The experimental data are compared with the 
theoretical results at $x$ shown by the vertical 
dotted line. 
The inset shows the schematic picture of the orbital 
structure 
below $T_{OO}$ at $x=0.125$. 
[$(\theta_A/\theta_B=-\theta_A)$ where $\theta_A=\pi/2$].
The
parameter values are chosen to be 
$J_1/t_0=0.25$, $J_2/t_0=0.0625$, and $J_{AF}/t_0=0.004$.}
\label{fig3}
\end{figure}
%
%

The mean field approximation is adopted in the calculation of the spin and
orbital states  at finite $x$ and $T$ \cite{degennes}.
Two kinds of the mean field for $\vec S_i$ and  $\vec T_i$ are introduced. 
The both states are described by the distribution functions and the mean
fields are optimized by minimizing the free energy. 
The ferromagnetic spin and G-type pseudo-spin  alignments are assumed. 
The detailed formulation will be presented elsewhere. 

The calculated phase diagram is presented in Fig.~\ref{fig3}. 
With doping of holes, a leading magnetic interaction gradually turns from 
the SE interaction in the lower $x$ to the DE one \cite{maezono}. 
$T_C$ monotonically increases with increasing $x$. 
On the other hand,  the orbital state changes from 
the AF-type ordering favored by ${\cal H}_{J}$ to the ferro-type ordering
induced by ${\cal H}_t$ due to the gain of the kinetic energy. 
Thus, $T_{OO}$ decreases with doping of holes. 
In the undoped insulator,   
$T_{OO}$ is higher than $T_C$ because the interaction between 
orbitals $(3J_1/2)$ are
larger than that between spins $(-J_1/2)$, 
as expected from the first term in ${\cal H}_{J}$.  
Consequently, 
$T_C$ and $T_{OO}$ cross with each other at $x_c \sim 0.1$ as seen in
Fig.~\ref{fig3}.

\begin{figure}
\epsfxsize=6.5cm
\centerline{\epsffile{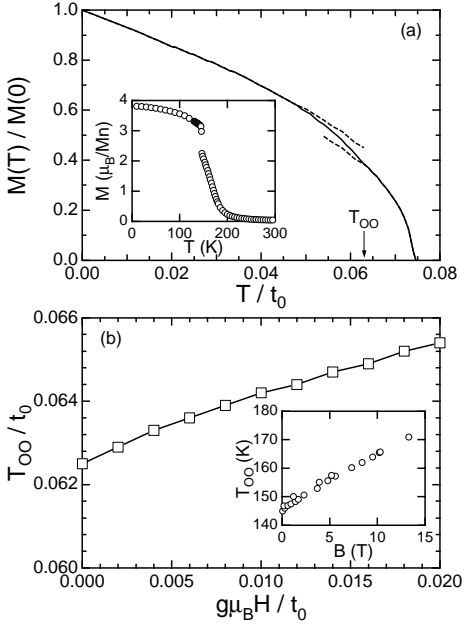}}
%
%
\caption{(a) Temperature dependence of the calculated
magnetization at $x=0.125$.  
The inset shows the magnetization 
curve in $\rm
La_{0.88}Sr_{0.12}MnO_3$ at $B=0.5T$. 
(b) Field dependence of the orbital ordering temperature. The inset
is obtained from the electrical resistivity and the magnetization in $\rm
La_{0.88}Sr_{0.12}MnO_3$ \protect\cite{nojiri}. 
Parameter values are the same with those in
Fig.~\protect\ref{fig3}.}
\label{fig4}
\end{figure}

We next focus on the region where $x$ is slightly higher than $x_c$. 
There are two kinds of the ferromagnetic phase; the phase
between $T_C$ and $T_{OO}$ and that below $T_{OO}$. 
In the high-temperature phase, the orbital is disordered. 
In the low-temperature phase, on the other hand, 
the AF-type orbital ordering appears and the SE 
interaction is enhanced through the spin-orbital coupling 
in $H_J$.
Since the AF-type ordering reduces the kinetic energy, 
the DE interaction is weakened. 
Consequently, the metallic character is degraded. 
We identify the low- and high-temperature 
phases to be FI and FM in $\rm La_{0.88}Sr_{0.12}MnO_3$, 
respectively.   
Note again that
the AF-type orbital ordering is driven by the electronic mechanism, not 
supported
by the JT distortion \cite{yunoki}.  
The orbital ordering (the inset of
Fig.~\ref{fig3}) is denoted  as
$(\theta_A/\theta_B=-\theta_A)$ with $\theta_A=\pi/2$,  
where $\theta_{A(B)}$ is the angle in the 
orbital space in the $A(B)$ sublattice. 
This is the mixture of 
$d_{z^2-x^2(y^2-z^2)}$ and $ d_{3x^2-r^2(3y^2-r^2)}$, 
which is consistent with the isotropic ferromagnetic 
spin wave dispersion. 
The characteristic curve in the azimuthal angle dependence  of the resonant
x-ray scattering (Fig.~\ref{fig2}(b)) is reproduced by this type of 
the ordering \cite{ishihara3}. 
The coupling between spin and orbital 
reflects on the temperature dependence of the magnetization.
It is shown in Fig.~\ref{fig4}(a), 
that the magnetization is enhanced below $T_{OO}$. 
The calculated result is consistent with the 
experimental observation in 
$\rm La_{0.88}Sr_{0.12}MnO_3$ (the inset of Fig.~\ref{fig4}(a)).  
This is a strong evidence of the novel coupling between  spin and
orbital degrees.

In Fig.~\ref{fig4}(b), $T_{OO}$ is shown as a function of the applied magnetic
field.  
The applied magnetic field stabilizes  the low-temperature  ferromagnetic
phase accompanied with the orbital ordering. This is because the
ferromagnetic spin correlation induced by the field  enhances 
the interaction between orbitals through  the spin-orbital coupling in ${\cal
H}_J$. In other word, the magnetic field controls the orbital states. 
The theoretical $T_{OO}$ versus $H$ curve qualitatively reproduces the experimental 
results 
in $\rm La_{0.88}Sr_{0.12}MnO_3$ (the inset 
of
Fig.~\ref{fig4}(b)) and strongly supports that the orbital degree
plays a key role  in the low temperature phase and the transition at $T_{
OO}$. 

%
%

To conclude, the transition from the ferromagnetic metallic to the
ferromagnetic insulating phases in $\rm La_{0.88}Sr_{0.12}MnO_3$ is ascribed
to the transition of orbital order-disorder states. 
The orbital ordering  is observed in the low temperature phase where the
cooperative Jahn-Teller type distortion is significantly diminished. 
The stability of the two phases are controlled by changing temperature  
and/or
applying magnetic field, and the unique coupling between  spin and orbital
degrees is found.  The present investigation shows a novel role of the 
orbital 
degree of freedom as a hidden parameter in the MI transition in lightly 
doped
CMR manganites. 

Authors acknowledge D. E. Cox, K. Tsuda, and T. Inami for their valuable discussions.  
Part of the numerical calculation was performed in the HITACS-3800/380 
superconputing facilities 
in IMR, Tohoku University. 
S.O. acknowledges the financial support of JSPS Research Fellowships 
for Young Scientists. 

\vfill

\begin{references}
%
\bibitem{zener}  C. Zener,  Phys.\ Rev.\ {\bf 82}, 403 (1951). 
%
\bibitem{anderson} P. W. Anderson, and H. Hasegawa,  Phys.\ Rev.\ {\bf 100},
 675
(1955). 
%
%
\bibitem{millis} A. J. Millis {\it et al.}, 
Phys.\ Rev.\ Lett. {\bf 77}, 175 (1996). 
%
\bibitem{urushibara} A. Urushibara {\it et al.}, Phys.\ Rev.\ B {\bf 51},
14103
(1995).  
%
\bibitem{kawano} H. Kawano {\it et al.}, Phys.\ Rev.\ B {\bf 53}, 
R14709 (1996). 
%
\bibitem{murakami} Y. Murakami {\it et al.},  Phys.\ Rev.\ Lett.\ {\bf 80}, 
1932
(1998), and Phys.\ Rev.\ Lett.\ {\bf 81}, 582 (1998). 
%
\bibitem{ishihara3} S. Ishihara and S. Maekawa,  Phys.\ Rev.\ Lett.\ {\bf 
80},
3799 (1998), and   Phys.\ Rev.\ B {\bf 58} 13449 (1998).
%
\bibitem{argyriou} D. N. Argyriou {\it et al.}, Phys.\ Rev.\ Lett. {\bf 76} 
3826
(1996).
%
%
\bibitem{senis} R. Senis {\it et al.},  
Phys.\ Rev.\ B \ {\bf 57},14680 (1998).
%
\bibitem{uhlenbruck} S. Uhlenbruck {\it et al.},  Phys.\ Rev.\ Lett.\ {\bf 
82},
185 (1999).
%
\bibitem{nojiri} H. Nojiri {\it et al.},  (unpublished). 
%
\bibitem{cox} D. E. Cox {\it et al.},  (unpublished). 
%
\bibitem{tsuda} K. Tsuda {\it et al.},  (unpublished).
%
\bibitem{yamada} Y. Yamada {\it et al.}, Phys.\ Rev.\ Lett.\ {\bf 77}, 904
(1996). 
%
\bibitem{inami} T. Inami {\it et al.},   (to be published in Jpn.\ J.\ Appl.
\
Phys.)  
%
\bibitem{fukuda} T. Fukuda {\it et al.}, (unpublished).
%
\bibitem{hirota1} K. Hirota {\it et al.},   Physica {\bf B 237-238}, 
36 (1997). 
%
\bibitem{hirota2} K. Hirota {\it et al.},  J.\ Phys.\ Soc.\ Jpn.\ {\bf 65}, 
3736
(1996).  See also F. Moussa {\it et al.}, Phys.\ Rev.\ B {\bf 54}, 15149 
(1996).
%
\bibitem{ishihara1} S. Ishihara {\it et al.},  Phys.\ Rev.\ B {\bf 55}, 8280
(1997). 
%
\bibitem{ham}
$t_{ij}^{\gamma \gamma'}$ is the transfer intensity between 
$i$ and $j$ sites with $\gamma$ and $\gamma'$ orbitals. 
$J_1=t_0^2/(U'-I)$ and $J_2=t_0^2/(U'+I+2J_H)$, where 
$t_0$ is the orbital independent transfer intensity, 
$U'$ is the Coulomb interaction between different orbitals  and $I$ is the
exchange interaction between $e_g$ electrons. 
$J_1 > J_2$ is satisfied. 
%
\bibitem{goodenough} J. B. Goodenough,  Phys.\ Rev.\ {\bf 100}, 564 (1955). 
 
%
\bibitem{kanamori} J. Kanamori,  J.\ Phys.\ Chem.\ Solids {\bf 10}, 87 
(1959).
%
\bibitem{kugel} K. I. Kugel, and D. I. Khomskii,  ZhETF Pis.\ Red.\ {\bf 15}
,
629 (1972), (JETP Lett.\ {\bf 15}, 446 (1972)). 
%
\bibitem{degennes}  P. G. de Gennes,  Phys.\ Rev.\ {\bf 118}, 141 (1960).
%
\bibitem{maezono}  R. Maezono {\it et al.},  Phys.\ Rev.\ B {\bf 57}, R13993
(1998).
%
\bibitem{yunoki} 
Two kinds of the ferromagnetic phase driven 
by the DE interaction and the JT distortion are 
discussed in 
S. Yunoki {\it et al.}, 
Phys.\ Rev.\ Lett.\ {\bf 81}, 5612 (1998).  
%
%
\end{references}
\end{document}